\documentclass[aps,reprint,amsmath,amssymb,eqsecnumaps,pra,superscriptaddress]{revtex4-2}
\usepackage{graphicx}
\usepackage{dcolumn}
\usepackage{bm}
\usepackage{float}
\usepackage{xcolor}
\usepackage{lineno}
\usepackage{verbatim} 
\usepackage[normalem]{ulem}
\usepackage{lineno}
 
\newcommand{\p}{\partial}

\newcommand{\vta}{\vartheta}
\newcommand{\kk}{k}
\newcommand{\bk}{K}
\newcommand{\bj}{J}
\newcommand{\om}{\omega}

\newcommand{\nn}{\nonumber}
\newcommand{\ta}{\theta}

\newcommand{\al}{\alpha}
\newcommand{\wt}{\tilde}
\newcommand{\pz}{\boldsymbol{\hat\sigma_z}}
\newcommand{\px}{\boldsymbol{\hat\sigma_x}}
\newcommand{\wh}{\widehat}

\newcommand{\cz}{{\cal Z}}

\newcommand{\cW}{{\cal W}}

\newcommand{\cA}{{\cal A}}

\newcommand{\be}{\begin{equation}}                                       
	\newcommand{\ee}{\end{equation}}
\newcommand{\ba}{\begin{eqnarray}}
	\newcommand{\ea}{\end{eqnarray}}
\newcommand{\bref}[1]{(\ref{#1})}

\newcommand{\bi}[1]{\bibitem{#1}}\newcommand{\lab}[1]{\label{#1}}

\newcommand{\bsub}{\begin{linenomath}\begin{subequations}}                      
		\newcommand{\esub}{\end{subequations}\end{linenomath}}     

\newcommand{\bnol}{\begin{linenomath}}                      
	\newcommand{\enol}{\end{linenomath}}     
\newcommand{\ket}[1]{\vert{#1}\rangle}
\newcommand{\bra}[1]{\langle{#1}\vert}

\begin{document}
	
	\preprint{APS/123-QED}
	
	\title{
Soliton metacrystals: topology and chirality
}
	\author{Z. Fan}
	\affiliation{Department of Physics, University of Bath, Bath BA2 7AY, UK}
\affiliation{Centre for Photonics and Photonic Materials, University of Bath, Bath, BA2 7AY, UK}%
	\author{D.N. Puzyrev}
	\affiliation{Department of Physics, University of Bath, Bath BA2 7AY, UK}
\affiliation{Centre for Photonics and Photonic Materials, University of Bath, Bath, BA2 7AY, UK}%
	\author{D.V. Skryabin}
	\email{d.v.skryabin@bath.ac.uk}
	\affiliation{Department of Physics, University of Bath, Bath BA2 7AY, UK}
	\affiliation{Centre for Photonics and Photonic Materials, University of Bath, Bath, BA2 7AY, UK}%
	
	\begin{abstract}
Designing metamaterials with the required band structure, topology and chirality using nano-fabrication technology revolutionises modern science and impacts daily life. The approach of this work is, however, different. We take a periodic sequence, i.e., metacrystal, of the dissipative optical solitons rotating in a single ring microresonator and demonstrate its properties as of the 
electromagnetic metamaterial acting in the radio to terahertz frequency range. The metacrystal unit cell consists of the bound pair of solitons, and the distance between them is used as a control parameter. We are reporting the soliton metacrystal band structure and its topological properties. The latter is confirmed by the existence of the $\pi$ steps experienced by the crystal phonons' geometrical (Zak) phase. Furthermore, we found the phononic edge states in the metacrystals with defects made by removing several solitons. Optical frequency combs corresponding to the soliton metacrystals reveal the spectral butterfly pattern serving as a signature of the spatio-temporal chirality and bearing a resemblance to the butterfly wings illustrating natural occurrences of chirality. 
	\end{abstract}
	\maketitle
	

Two prolific themes spanning across today's solid-state, cold-atom and optical physics are the topology of waves in periodic potentials~\cite{solid,cold,pc,book,nnb,nnc,mec1} and localised coherent states, such as, e.g., solitons and vortices. The recent and ongoing work on the vortex and skyrmion matter and light~\cite{bog,skyr,fo,su}, soliton crystals~\cite{prls,de,papp,kip,syn,don,tah,tahr} and soliton gas~\cite{gas1,gas2}  points at the opportunity of using the localised structures 
of the coherent light and matter waves for designing  new electromagnetic materials. 
Here, we present the proof-of-concept results revealing the topological properties of the periodic soliton sequences, i.e., soliton metacrystals, generated in optical microresonators.

For this work, we should distinguish the dissipative and conservative optical solitons and single out the former for their robustness and longevity~\cite{gr}. Crystals of dissipative optical solitons are known to exist in Kerr ring microresonators~\cite{papp,kip,don,tah,syn}, fibre lasers~\cite{cr0,cr1} and in the exciton-polariton resonators~\cite{krizh}. Methods of the on-demand positioning of the individual dissipative solitons, circling with the typical repetition rates between ten GHz and one THz, inside the ring resonators have also been successfully demonstrated. These methods included -  pump modulation~\cite{syn}, using bi-chromatic pump~\cite{tahr}, single pump frequency tuning \cite{kip0}, and 
the excitation of the desired soliton ordering by applying a sequence of pulses with the repetition 
rate higher than the resonator free spectral range~\cite{mir}. Though the soliton crystals in optical resonators are well established,  their band structure and its topology remain an uncharted territory. 

The focus of many prior studies of topological properties of nonlinear optical systems has been on the cases when taking the no-nonlinearity limit leaves a linear system prepossessing a specially designed linear periodic structure, see, e.g., \cite{sol0,sol1,sol2,sol3,mot} 
for topological solitons and \cite{p1,p2,p3,prl2,prl3,ssh2,se1,ley,chembo,yk} for other topology vs 
nonlinearity interplay in the arrays of optical resonators and waveguides, and photonic crystals. The soliton crystals studied below require only a single optical resonator and are sustained by the pump-loss and dispersion-nonlinearity balances. 

In this work, we demonstrate the soliton metacrystals consisting of the bound soliton pairs so that the soliton arrangement looks like the famous Su-Schrieffer-Heeger (SSH) lattice~\cite{book}. The relative positioning of the solitons within a unit cell is used as the control parameter helping to reveal topological and chiral properties. To prove the non-trivial topology of the metacrystal band structure, we develop a theory of their geometrical phase~\cite{ber,zak2}, and also demonstrate the topological edge states~\cite{book} in the metacrystals with defects. The edge states typically come in chiral pairs as, e.g., the edge currents on the opposite interfaces of the suitably cut two-dimensional topological crystal~\cite{pc,book}. A feature of the multimode resonators is that the trains of the modelocked solitons they generate are periodic in space and time. We found that the spatio-temporal chirality of the soliton metacrystals leads to the distinct butterfly-like structure hidden in their optical spectra. Various practical and conceptual realizations of time crystals in condensed matter and optics~\cite{wil,t1,t2,t2a,t3,t4,t5} and the cross-disciplinary applications of chirality~\cite{pen2,chir2,ch9,wil2} are currently attracting the increasing attention.

\begin{figure*}[t]
	\centering{	
		\includegraphics[width=0.95\textwidth]{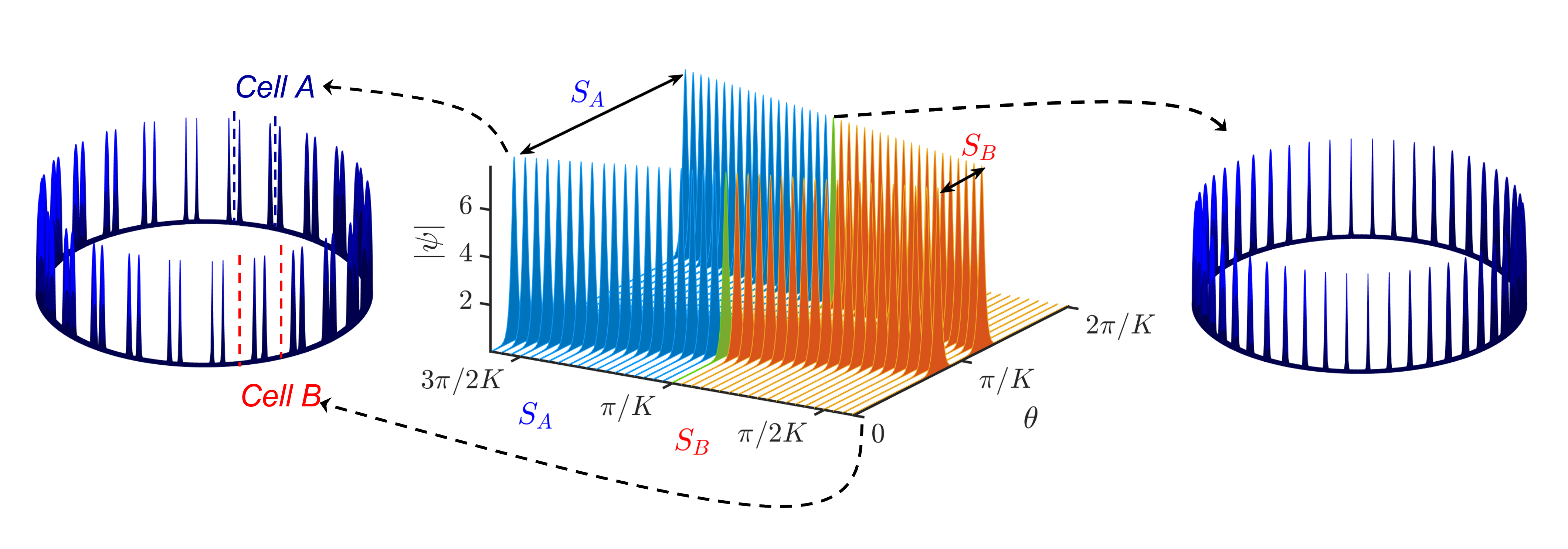}	
	}
	\caption{{\bf Soliton metacrystals and dimerization choices.} 
		The middle panel shows how two solitons are positioned in the unit cell for varying separation distance, $S_\al$. Two side panels show all $2K=48$ solitons along the ring circumference. The right plot corresponds to $S_A=S_B$, i.e., when the molecular crystal degenerates to the atomic one. The left plot shows the molecular metacrystal and illustrates 
		the two dimerization choices applied in this work, $S_A+S_B=2\pi/K$.
	   The scaled dispersion and detuning parameters  are    $d_2/\kappa=5\times 10^{-4}$,  $d_3/\kappa=5\times 10^{-7}$, and $\delta/\kappa=30$. The crystal repetition rates vary with $S_\al$, but $d_1/\kappa$ remains close to $-0.01059$. The dimensionless field amplitude, $\psi$, is expressed as $\gamma |\phi|^2/\kappa=|\psi|^2$. The dimensionless pump parameter is $\gamma h^2/\kappa=25$. Crystals are computed with $512$ points across the unit cell,  $\ta\in[0,2\pi/K)$.  }
	\label{f1}
\end{figure*} 

\section*{RESULTS AND DISCUSSION}
\subsection*{Band structure of soliton metacrystals}
We consider an optical ring microresonator~\cite{papp,kip,syn,don} which frequency spectrum is approximated by $\om_\mu=\om_0+D_1 \mu+D_2 \mu^2/2!+D_3 \mu^3/3!$, where $\mu=0,\pm 1,\pm 2,\dots$ is the mode number, $D_1$ is the linear repetition rate (inverse of the round-trip time), and $D_{2,3}$ are the second and third order dispersion coefficients. 
$\om_\mu$ typically belong to the infrared-to-visible range.
The pump laser frequency $\om_p$ is tuned around $\om_0$, so that $\delta=\om_0-\om_p$ is the detuning parameter. 
The envelope,~$\phi$, of the multimode optical field inside the resonator is 
\be
\phi(\ta,t)=\sum_{\mu}\phi_\mu(t) e^{i\mu\ta}=\phi(\ta+2\pi,t),~\ta=\vta-\wt D_1 \tau,
\lab{e2}\ee
where, $\phi_\mu$ are the mode amplitudes, $\vta\in [0,2\pi)$ is the polar angle, and $\wt D_1$ is the nonlinear repetition rate. 

A soliton crystal, with the spatial period $2\pi/K$ ($K=1,2,\dots$) and the
temporal one $2\pi/K\wt D_1$, is defined as a particular realisation of Eq.~\bref{e2},  
\be
\phi(\ta,t)=\Phi(\ta)=\sum_{l}\phi_{l\bk} e^{il\bk \ta}=\Phi\left(\ta+\frac{2\pi}{\bk}\right),
\lab{e3}
\ee 
where the non-zero modes are separated by the step $\bk$, $\mu=l\bk $ 
($l=0, \pm 1,\pm 2,\dots$). The corresponding optical spectrum of the soliton crystal is equidistant and is expressed as  $\wt\om_{lK}=\om_p+lK\wt D_1$. The envelope $\phi$ obeys the microresonator implementation of the Lugiato-Lefever model~\cite{a1}, 
\be
i\p_t\phi =\delta\phi
+
\sum_{j= 1}^3\frac{d_j}{j!}(-i\p_\ta)^j \phi
-\gamma |\phi|^2\phi
-\frac{i\kappa}{2}(\phi-h),
\lab{ll0}
\ee
where  $d_1=D_1-\wt D_1$, $d_{2,3}=D_{2,3}$, $\kappa$ is the linewidth, $h^2$ is the 
pump power, and $\gamma$ is the nonlinear parameter. 

If a single soliton and its small excitations correspond to an atom, 
then a bound state of two solitons is a soliton-molecule~\cite{mol2,mol3,mol5,mol7}. 
A periodic arrangement of the soliton-molecules makes up what we call a metacrystal. 
The numerically computed family of such crystals with 
two solitons per unit cell, $\ta\in[0,2\pi/K)$, and $K=24$  is shown in Fig.~\ref{f1}.  
The metacrystal structure is  SSH-like~\cite{book}, but 
none of the terms in Eq.~\bref{ll0} is modulated, i.e., our  
lattice is  self-sustained by the dispersion-nonlinearity 
and pump-loss balances. When the inter-soliton distance equals half of the unit cell length, $\pi/K$,
then the metacrystal (molecular crystal) with period $2\pi/K$ degenerates 
to the usual (atomic) soliton crystal~\cite{papp,kip}.

The difference of the interaction strengths between the left and right neighbours of a given soliton, in other words,  the imbalance of the intra-cell and inter-cell coupling rates, is controlled by the separation distance $S_\al$. There exist two choices of $S_\al$, i.e., dimerization choices, corresponding to the same crystal,  $S_\al=S_A$ and~$S_\al=S_B=2\pi/K-S_A$ (see Fig.~\ref{f1}).
In what follows, we take $\pi/K\leqslant S_A< 2\pi/K$ and $0< S_B\leqslant\pi/K$, so that the choice of the unit-cell~$A$, i.e., dimerization $A$,  corresponds to the inter-cell coupling  being stronger than the intra-cell one. $S_{A,B}=\pi/K$ is the degeneracy point with  equal coupling rates.

Soliton metacrystals should modify the spectrum of small amplitude perturbations, i.e., elementary excitations, supported by the resonator. Physically, these excitations are slow modulations of the amplitudes and phases of the resonator modes having  frequencies in the radio- to tera-hertz range. Using an analogy with the density waves in the solid state crystals, we term these excitations - phonons. While the nonlinear interactions between the crystal and phonon modes are imperative in what follows,  phonon to phonon interactions are weak and should be disregarded.

\begin{figure*}[t]
	\centering{		\includegraphics[width=0.9\textwidth]{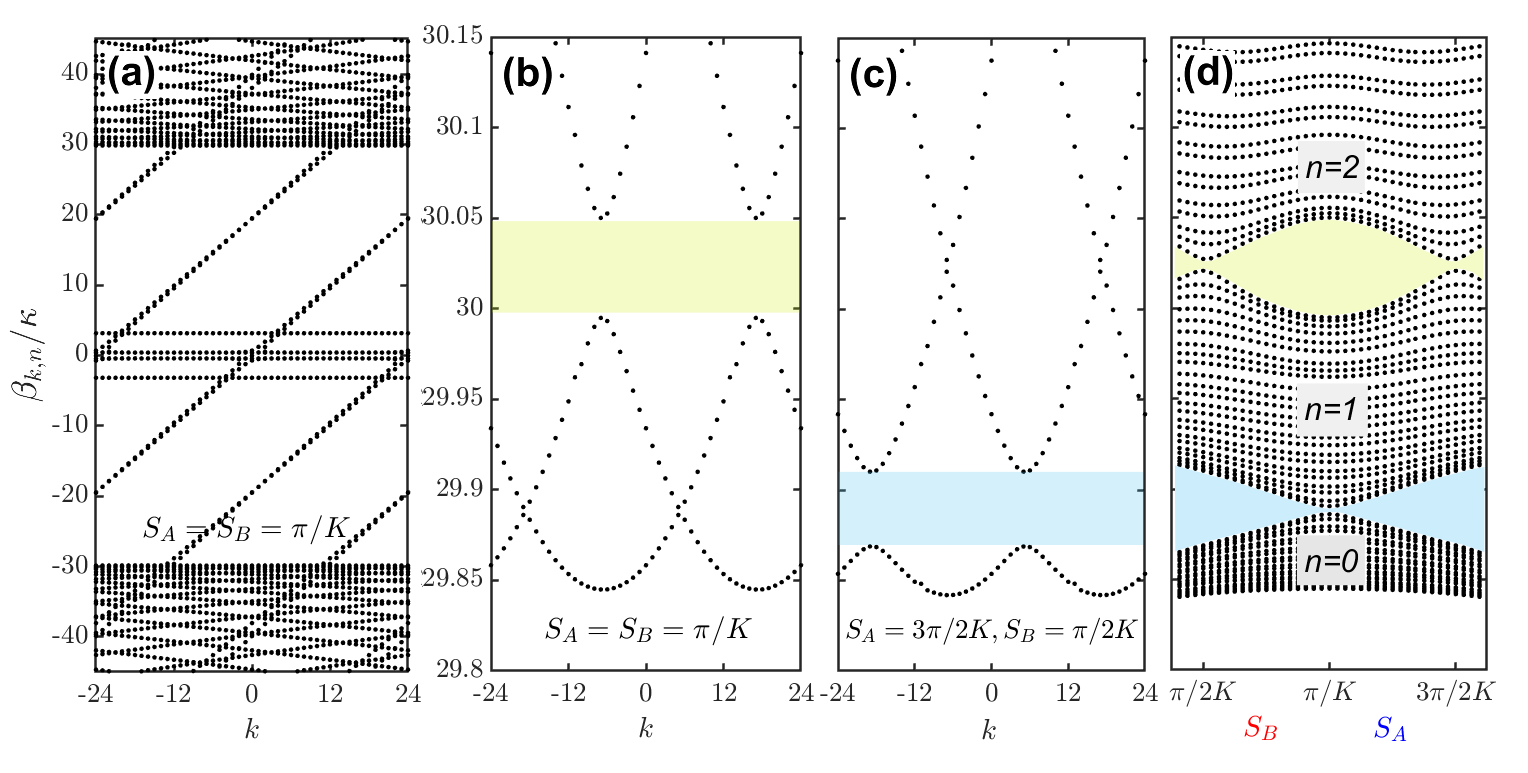}	
	}
	\caption{{\bf Phononic band structure in soliton metacrystals.} Panels (a,b,c) show the spectrum vs  momentum $k$ varying across the two Brillouin zones. 
The number of Bloch states in one Brillouin zone is $K=24$. 
		Panels (b,c) zoom on the band structure around the top edge of the continuum, see $\beta_{k,n}/\kappa\approx 30$ in (a).
 Panel (d) demonstrates how the band structure changes with the inter-soliton distance $S_{\al}$. 
 		The  blue and yellow  shading in (b,c,d) highlight the first and second bandgaps.  
}	
	\label{f2}
\end{figure*}

The phonon band structure is computed using the Bloch theorem. This is achieved by taking a soliton crystal, Eq.~\bref{e3}, and its complex conjugate, and representing the phonon field  by the cell-periodic two-component Bloch vectors, 
${\bf u}^{(\al)}_{k,n}(\ta)={\bf u}^{(\al)}_{k,n}(\ta+2\pi/K)$, 
\ba
\begin{bmatrix}
	\phi\\
	\phi^*
	\end{bmatrix}=&&
\begin{bmatrix}
	\Phi^{(\al)}\\
	\Phi^{(\al)*}
\end{bmatrix}+\sum_{k,n}e^{t\lambda_{k,n}}\times\left[{{\bf u}^{(\al)}_{k,n}}e^{ik\ta-it \beta_{k,n}}\right.\nn
\\
&&
~~~\left.+\px{{\bf u}^{(\al)*}_{k,n}}e^{-ik\ta+it \beta_{k,n}}\right],\px=
\begin{bmatrix}
	0 & 1\\
	1 & 0
	\end{bmatrix}.
\lab{f7}
\ea
Every Bloch vector is characterised by  (i)~Bloch momentum,  $k$, which
can be restricted to the first Brillouin zone, $k=1,2,\dots,K$, (ii)~band index, $n$, and  (iii)~dimerization index, $\al=A,B$. 
$\beta_{k,n}=\beta_{k+K,n}$ are 
the phonon frequencies and   $\lambda_{k,n}=\lambda_{k+K,n}$ express 
the balance between the dissipation  and  parametric gain. 
The momentum varies discretely because of the  ring geometry.
The condition of the crystal stability, $\lambda_{k,n}\leqslant 0$, 
is satisfied for all solutions shown in Fig.~\ref{f1}.
Below we use the  ket notation for ${\bf u}^{(\al)}_{k,n}=\ket{{\bf u}^{(\al)}_{k,n}}$ and 
bra for its transpose and complex conjugate, $\px{\bf u}^{(\al)*}_{k,n}=\bra{{\bf u}^{(\al)}_{k,n}}$. 

The Bloch states solve the eigenvalue problem
\be
\pz\wh H_k^{(\al)}\ket{{\bf u}^{(\al)}_{k,n}}
=
(\beta_{k,n}+i\lambda_{k,n}+i\tfrac{1}{2}\kappa)\ket{{\bf u}^{(\al)}_{k,n}},
\pz=
\begin{bmatrix}
	1 & 0\\
	0 & \text{-}1
\end{bmatrix},
\lab{jb}\ee
where $\wh H_{k}^{(\al)}$  is the system Hamiltonian, see  
Methods for details.
The phonon spectrum, unlike the eigenstates, does not change if $\al=A$ is replaced with $B$.
Fig.~\ref{f2}a shows the phonon spectra of the soliton crystals with $S_A=S_B$. The values of $\beta_{k,n}$ scale with the linewidth, $\kappa$, which, in the chip-integrated microresonators, varies in the range from tens to hundreds of MHz. The continuum of states starts around $\beta_{k,n}\approx\pm\delta$ and is shaped like a typical band structure. Fig.~\ref{f2}b narrows the view to the top edge of the continuum, where one can see the band-crossing feature and bandgap (yellow shading). $d_3\ne 0$ produces the small amplitude dispersive waves~\cite{cr2} with the strongly tilted straight-line spectrum, see Fig.~\ref{f2}a.

\begin{figure*}[hptb]
	\centering{\includegraphics[width=0.9\textwidth]{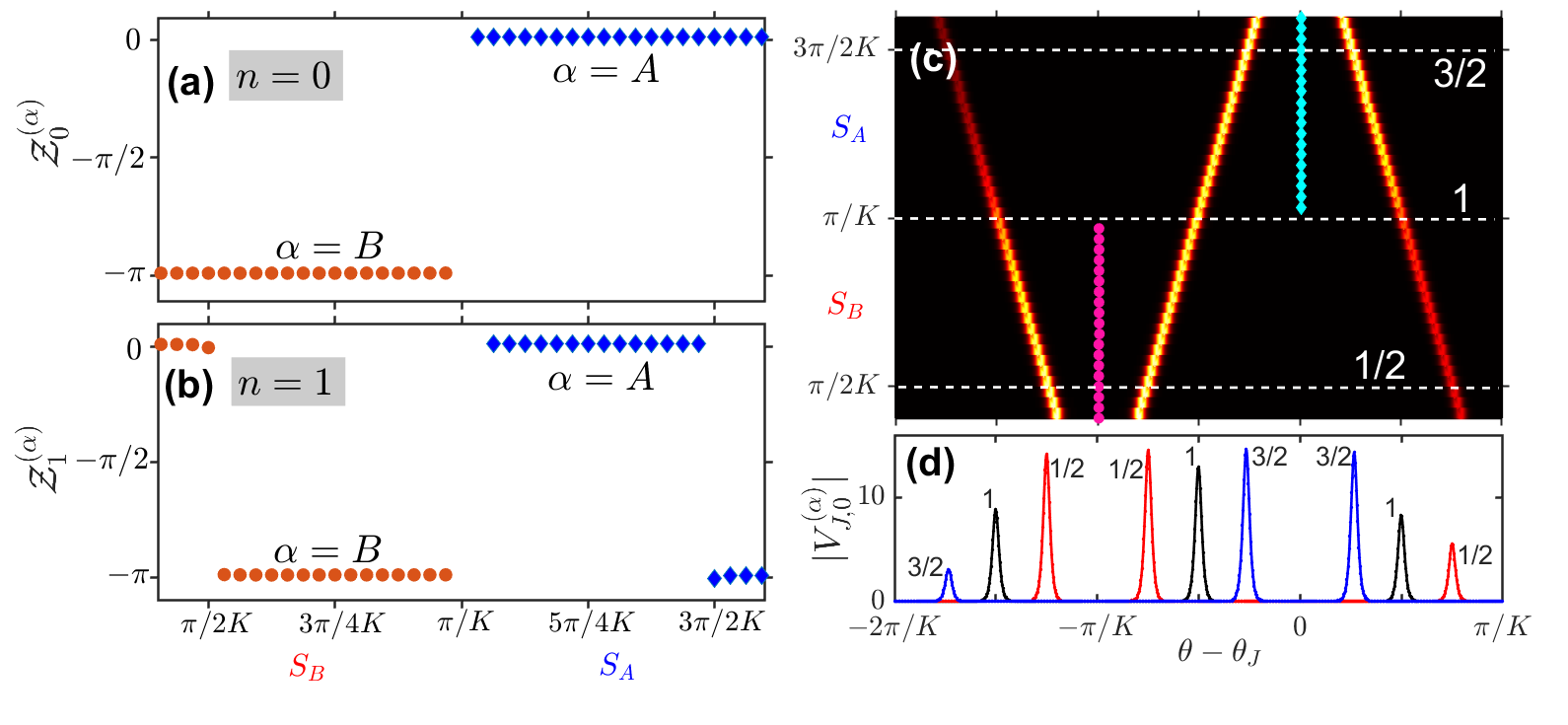}	
	}
	\caption{{\bf Geometrical (Zak) phases and Wannier functions in soliton metacrystals.}  (a,b)~Numerically computed Zak phases for the $n=0$ (a) and $n=1$ (b) bands. The phases flip by $\pi$ at the band-crossing points, cf.~Fig.~\ref{f2}d. (c)~The change of the positions and relative strength of the peaks of the $n=0$ Wannier function vs  the soliton separation, $S_\al$, and $\ta-\ta_J$ (restricted to the one 	and a half unit cell length to boost resolution). The black-orange colour-map shows $|V_{J,0}^{(\al)}|$, see Eq. \bref{b7}. The magenta ($\al=B$) and blue ($\al=A$) vertical doted lines mark the values of $\cz^{(\al)}_0/K$ from (a). (d)~shows three selected profiles of $|V_{J,0}^{(\al)}|$ for $S_\al$ marked by the white horizontal lines in (c).  }	
	\lab{f3}
\end{figure*}

When $S_A$ becomes $\ne S_B$, either way, the band crossing opens up to the bandgap, see the blue shading in Fig.~\ref{f2}c. At the same time, the yellow shaded $S_A=S_B$ band gap starts narrowing and comes to the near band-crossings for $S_{A,B}= \pi (1\pm 0.5)/K$, see Fig.~\ref{f2}d. We recall that the textbook SSH chain with the nearest-neighbour coupling has only two bands that cross at the Brillouin zone edges when the inter and intra-cell coupling rates ($r$ and $r'$) are equal. 
The topological gap opens up for $r\ne r'$, and its width equals  $|r-r'|$~\cite{book}. Thus, the first two bands in the soliton crystal lattice behave very similarly to what happens in the SSH  chain. 
To some extent, the similarity with the SSH model comes unexpectedly, since the spectrum of the Jacobian operator, $\pz\wh H_{k}^{(\al)}-i\tfrac{1}{2}\kappa\mathbf{\hat 1}$, 
defines the phonon band structure in soliton crystals, while in the SSH model~\cite{book}, and for the non-interacting electrons in crystals~\cite{zak2}, this is the Hamiltonian spectrum.


\subsection*{Geometrical phase and Wannier-centres}
If the soliton crystal is considered from the same point of view as the ionic crystal lattice shaping clouds of bound electrons in dielectrics, then the next natural step would be to understand the shapes of the phononic clouds existing around the solitons. This is achieved by developing the theory like the theory of polarization in crystalline dielectrics, where it has also provided a transparent interpretation of the geometrical (Zak) phase of an electron acquired after its passage over the Brillouin zone (BZ)~\cite{zak1,zak2,resta,van1}. 

The $k$-space periodicity and gauge-freedom properties of the Bloch vectors are,
\be
{\bf u}_{k,n}^{(\al)}={\bf u}_{k+K,n}^{(\al)}e^{iK\ta},~ 
{\bf u}_{k,n}^{(\al)}\to {\bf u}_{k,n}^{(\al)}e^{i\xi_k}.\lab{ye1}
\ee
Here, $\xi_k$ are the  phases, which are arbitrary across the first  BZ and  extendable beyond it with $\xi_k=\xi_{k+K}$. 
The discreteness of $k$ requires special care when formulating the gauge invariant 
computational strategy for the geometric phase. 
Adopting the approach of~\cite{van1,resta}, we have derived  the  following equation for the geometric phase,  
\be
\cz^{(\al)}_n=\text{Im}\ln\prod_{k=1}^{K} \bra{ {\bf u}^{(\al)}_{k,n}}\pz \ket{{\bf u}^{(\al)}_{k+1,n}},
\lab{pro}
\ee
of phonons belonging to a specific band in the soliton crystals, see Methods. The simultaneous bra and ket occurrences for every~$k$ make Eq.~\bref{pro} gauge invariant and applicable to
any Bloch-function phases generated by the eigenvalue solver. 

We check  topological properties of phonons in soliton crystals numerically by observing the $\pi$-flipping of their geometric phase on dragging the system through the band crossing points. Figs.~\ref{f3}a,b show  $\cz_n^{(\al)}$ vs $S_\al$ for $n=0,1$. The $n=0$ phase flips from  $-\pi$ to zero  at the point $S_{A,B}=\pi/K$, where  the $n=0$ and $n=1$ bands make the  near-crossing, see Fig.~\ref{f2}d.  The topology of the $n=1$ band reflects on the co-existence of one crossing with the $n=0$ band and 
two crossings with the $n=2$ band, and, hence, its geometrical phase flips three times. 
 
\begin{figure*}[hptb]
	\centering{	
		\includegraphics[width=0.95\textwidth]{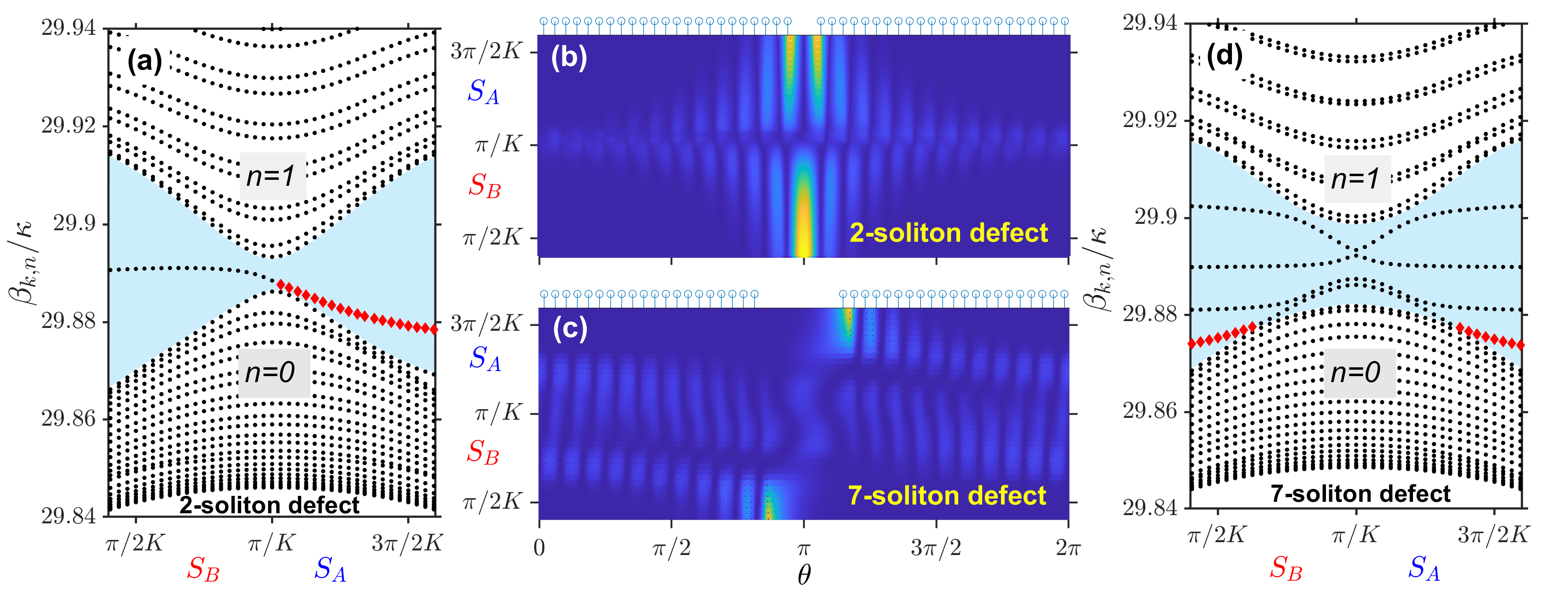}	
	}
	\caption{{\bf Phonon edge states in soliton metacrystals with defects.} (a) The phonon spectrum in and around the $n=0,1$ bandgap in the metacrystal with the defect made by removing one unit cell, i.e., 2-soliton defect.  	(b) The eigenfunctions of the bandgap states, $|{\bf u}_{k,n}^{(\al)}|^2$, 
		  vs $S_\al$ and $\ta$. The stems along top axes mark locations of the individual solitons and hence show the size of defects for $S_{A,B}=\pi/K$.
		  (c) is like (b), and (d) is like (a), but for the 7-soliton defect created by removing the three unit cells plus one soliton. 
		The red diamonds in (a) and (d) mark the spectra of the edge states.  Crystals with defects are computed with $16384$ points across the resonator circumference, $\ta\in[0,2\pi)$. $\ket{{\bf u}_{k,n}^{(\al)}}$ is computed 
		using Eq.~\bref{jb} with  $k=0$, $K=1$, which corresponds to $\ta\in[0,2\pi)$, see Methods for details.}
	\lab{f4}
\end{figure*}

If the number of cells is taken to infinity while their length is kept fixed at $2\pi/K$, then  
$k$ becomes continuous and, hence, $\p_k$ is well defined. In this limit, the geometric phase  acquires a more familiar form,  $\cz^{(\al)}_n= \int_0^K\cA^{(\al)}_{\kk,n}dk$, where $\cA^{(\al)}_{\kk,n}=i\bra{ {\bf u}^{(\al)}_{\kk,n}}\pz\p_\kk \ket{{\bf u}^{(\al)}_{\kk,n}}$. $\cA^{(\al)}_{\kk,n}$ 
differs from the Berry connection for the non-interacting electrons~\cite{zak2,ber} by the Pauli matrix $\pz$ before $\p_k$. Taking a unit cell 
with the coordinate $\ta=\ta_J$ and defining the corresponding  Wannier function as $\ket{{\bf w}_{J,n}^{(\al)}}=K^{-1}
\sum_{k} \ket{{\bf u}_{k,n}^{(\al)}}e^{ik(\ta-\ta_J)}$,  
we have further demonstrated that the geometric phase is directly proportional to 
the averaged coordinate of the phononic cloud centre in a given cell,
\be
\frac{\cz^{(\al)}_n}{K}= \bra{ {\bf w}^{(\al)}_{J,n}}\pz(\ta-\ta_J)\ket{{\bf w}^{(\al)}_{J,n}}.
\lab{z0}
\ee 
With $\pz\to\mathbf{\hat 1}$,  the above would be the equation for the
band- or Wannier-centre of the electron wave function in a crystal~\cite{zak2,zak1,resta,van1}, 
see Methods for more details.
For our unit cell definition, see Fig.~\ref{f1}, 
the $n=0$  phonon Wannier centres can be located either in the middle 
or at the boundary of a unit cell. 
Figs.~\ref{f3}c,d are explicit about the matching between $\cz^{(\al)}_0/K$ 
and  the geometric centres of the Wannier functions. 
Recent papers demonstrating measurements of the geometric phases of cold atoms in optical lattices~\cite{ssh0} and light in waveguide arrays~\cite{pes,long} have applied 
various practical methods to detect positions of the respective wave-packets. 

\subsection*{Metacrystals with defects:  Edge states and chirality}
For the SSH model, creating the lattice defect by removing one $A$-type unit cell, i.e., breaking two strong bonds in the chain, introduces the edge state inside the bandgap~\cite{book}. 
To study the bandgap states in the soliton metacrystals, we first extracted one pair of solitons and then tuned $S_\al$. $\al=B$ corresponds to breaking the weak bond and $\al=A$ to the strong one, see Fig.~\ref{f1}. 
The corresponding phonon spectrum plotted as a function of $S_\al$, see Fig.~\ref{f4}a, shows the near-crossing of the two bands at $S_\al=\pi/K$, and the state inside the bandgap emerging for $S_\al$ either $<\pi/K$ ($B$-type defect) or $>\pi/K$ ($A$-type defect).  The eigenvectors of the bandgap state  
are localised around the defect on either side from $\pi/K$, see Fig. \ref{f4}b. 
For $\al=A$, the state vector is localised on the edge solitons, while,
for $\al=B$, it is guided in the middle of the defect, see Fig.~\ref{f4}b.
The data presentation for the SSH edge states similar to the style chosen in Figs.~\ref{f4}b,c
can be found in~\cite{t3}.

If the defect is made by the odd number of solitons,  then shifting between the $\al=A$ and $\al=B$ dimerizations corresponds to moving the weak bond from one side of the defect to the other, see, e.g., the structure of the 7-soliton defect in Fig.~\ref{f5}a. Then, the edge state also migrates together with the weak bond, see Fig.~\ref{f4}c. With the defects larger than one unit cell, the spectrum in the bandgap contains not only the edge state but also other states guided inside the defect not at its edges, see the black doted lines inside the blue shading in Fig.~\ref{f4}d. The guidance mechanism for these modes is akin to the band-gap guidance in the hollow-core photonic crystal fibres~\cite{bc}.

\begin{figure*}[hptb]
	\centering{	
		\includegraphics[width=0.98\textwidth]{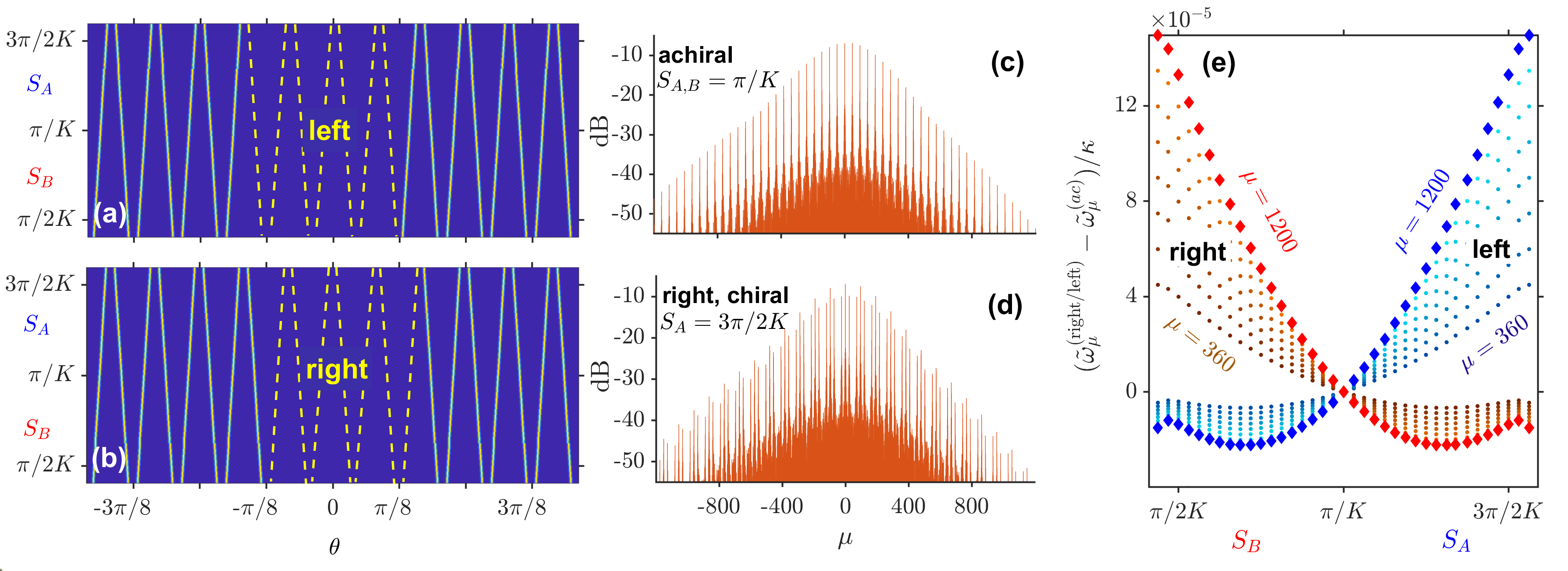}	
	}
	\caption{ {\bf  Chirality of soliton metacrystals with defects.} (a,b) show  metacrystals with the differently arranged 7-soliton defects and $K=24$. (a) and (b) geometries are called left and right, respectively. The missing soliton locations are marked with  dashed lines. Note that the $\ta$-interval shown is $<2\pi$. (c,d) show the examples of spectra of the achiral and chiral defects. (e) shows the normalised  frequency spectra of the crystals with the right  and  left defects, i.e., $\wt\om_\mu^\text{(right)}$ and $\wt\om_\mu^\text{(left)}$, minus the $\wt\om_\mu^\text{(ac)}$ spectrum corresponding to the achiral geometry,  $S_\al=\pi/K$.  
Red dots correspond to the right defect, 
$(\wt\om_\mu^\text{(right)}-\wt\om_\mu^\text{(ac)})/\kappa$, and  blue dots to the left one,
$(\wt\om_\mu^\text{(left)}-\wt\om_\mu^\text{(ac)})/\kappa$.  	The values of $\mu$ shown are restricted to $\mu\in [360,1200]$, with the step being 120. }
	\lab{f5}
\end{figure*}

The defects consisting of the odd-number of solitons break the left-right symmetry of the metacrystals and thereby appear as an interesting object to study the spatio-temporal and spectral manifestations of chirality~\cite{wil2}. 
We consider a crystal to be chiral, providing that the mirror image of the solitons around the defects can not be made to coincide with the original soliton arrangement. This is true for any $S_A\ne S_B$, while, the $S_A= S_B=\pi/K$ case corresponds to the achiral defect. 
Indeed, the unpaired soliton can be removed either on the left or right side of the defect, and then the two crystals with the same $S_\al$ are connected by the chiral symmetry transformation. Tuning $S_\al$ and plotting the crystals around the defect edges makes the chirality obvious, see the tilts of the solitons at the defect edges in Figs.~\ref{f5}a and \ref{f5}b.

The chiral pairs of the right and left defects have the spectra which are hard to distinguish. Therefore, we have just shown one of them in Fig.~\ref{f5}d. The spectrum of the ahiral defect is shown for comparison in 
Fig.~\ref{f5}c.   The envelope of the large amplitude spectral lines match well with the spectra of the ideal crystals, while the small amplitude dense spectrum originates from the defect. To make the spectral signature of the chirality effect obvious,
we have taken the frequency spectra of the left and right defects, i.e.,  $\wt\om_\mu^\text{(left)}$ and
$\wt\om_\mu^\text{(right)}$, and subtracted from them the achiral spectrum, $\wt\om_\mu^\text{(ac)}$.
Here, $\wt\om_\mu^\text{(..)}=\om_p+\mu \wt D_1^\text{(..)}$ is the mode-locked (dispersion free) frequency spectrum of the soliton crystal with defect and $\wt D_1^\text{(..)}$ is its numerically computed free spectral range. $\wt D_1^\text{(..)}$ is a function of $S_\al$, pump and all the other parameters. We shall recall that the  mode-locked spectrum for the ideal crystal, $\mu=lK$, has been already defined after Eq.~\bref{jb}.
Fig.~\ref{f5}e shows how $(\wt\om_\mu^\text{(left)}-\wt\om_\mu^\text{(ac)})/\kappa$ and 
$(\wt\om_\mu^\text{(right)}-\wt\om_\mu^\text{(ac)})/\kappa$ vary with $S_\al$. The spectral differences  make the butterfly pattern with the symmetry axis at $S_{\al}=\pi/K$ and, thereby, reveal the spatio-temporal chirality of the soliton metacrystals.

\subsection*{Discussion}
We have conjectured and confirmed numerically that the train of solitons in microresonators can be considered as a spatio-temporal topological metamaterial, i.e., soliton metacrystals. The optical spectra of such crystals correspond to the octave wide frequency combs.
The elementary excitations, i.e., phonons, in metacrystals are the radio to terahertz modulations propagating across the spectrum of the resonator modes. We have computed the phononic Bloch states and their geometrical (Zak) phases. The Zak phase has also been expressed via the Wannier functions and demonstrated to have the meaning of the effective polarization of the phononic clouds around the solitons. 

Zak phase changes in the steps of  $\pi$ with the tuning of the soliton separation, which proves the non-trivial topology of the metacrystal band structure.  Further to this point, we have found the phononic edge states upon introducing the 
one unit cell and larger defects. We have demonstrated that the optical (photon) spectra of metacrystals with the left and right defects can be used to characterize chirality in time. Subtracting the optical spectrum of the achiral crystal from the chiral ones reveals the butterfly structure of frequencies, see Fig.~\ref{f5}e, similar to the butterfly wings illustrating chirality in nature.

Features of the proposed realisation of the SSH-like soliton lattice are in its spatio-temporal and dissipative nature and that it does not require spatial or temporal modulations of the material refractive index and pump laser and fabrication of the complex arrays of coupled resonators and waveguides. Further, we should note that one out of twenty microresonator soliton crystals illustrated in~\cite{papp} shows the SSH-like arrangement of soliton molecules with several embedded defects, see Fig. 3j. This suggests that the frequency combs with topological properties may have already been unintentionally captured in the experimental measurements. Our results intend to motivate further theory,  experiments and applications so that this topic can grow and take its place in the interdisciplinary areas of topological and ultrafast physics. We are also forecasting applications of the soliton metacrystals and their topological states in classical and quantum information processing, where the microresonator solitons and topological photonics are already making an impressive start~\cite{rf1,quant2,w4,w1,to1}.


\renewcommand{\theequation}{m.\arabic{equation}}
\vspace{8mm}
\setcounter{equation}{0}
\noindent{\bf METHODS}\vspace{2mm}\newline
\noindent In order to shorten  notations  we have dropped the band index, $n$, and the dimerization superscript, $(\al)$, in the first part of Methods.
They are subsequently reintroduced in the second and third parts.
\vspace{2mm}\\
\noindent{\bf Bloch formalism}\vspace{1mm}\\
To analyse and solve Eq.~\bref{e2}
we set the ansatz consisting of the 
time independent solution which 
could be a crystal with or without defect, $\Phi(\ta)=\Phi(\ta+2\pi)$,
plus the small amplitude phonon field, $\wt\phi(\ta,t)$. Hence,
\be
\phi=\Phi(\ta)
+ \wt\phi(\ta,t),~
\wt\phi(\ta,t)=\wt\phi(\ta+2\pi,t).
\lab{l2}
\ee
Linearising  Eq.~\bref{e2}  for $\wt\phi$ yields
\be
	i(\p_t+\tfrac{1}{2}\kappa)\begin{bmatrix}
		\wt\phi \\ \wt\phi^*
	\end{bmatrix}=\pz\wh H_0\begin{bmatrix}
	\wt\phi \\ \wt\phi^*
\end{bmatrix}, \ta\in[0,2\pi),
\lab{xs1}
\ee
where $\wh H_0$ is specified in Eq.~\bref{ub}.

We now assume that $\Phi(\ta)$ has a period  
$2\pi/K$, where $K$ could also be one,  
\be
\Phi(\ta)=\sum_{l}\phi_{l\bk} e^{il\bk \ta}=\Phi\left(\ta+\frac{2\pi}{\bk}\right).
\lab{l3}
\ee 
Equations for $\phi_{l\bk}$ have been solved using the Newton method. 
Parameters used are typical  for Si$_3$N$_4$ resonators:
$D_1/2\pi=100$GHz, $D_2/2\pi=50$kHz, $D_3/2\pi=50$Hz, 
$\gamma/2\pi=30$MHz/W, $\kappa/2\pi=100$MHz 
and  laser power $\cW=520$mW, so that $h^2=(D_1/2\pi\kappa)\cW=83$W, 
see Eqs.~\bref{e2}-\bref{ll0} and Ref.~\cite{a1} 
for further parameter and model discussion. 
The scaled  values of the parameters are given 
in the Fig.~\ref{f1} caption. The approximate analytical expressions for the non-topological  
soliton crystals in the Lugiato-Lefever equation 
have been reported in, 
e.g.,~\cite{cam1,cam2}.

The phonon field is  set as~\cite{a4,a3} 
\ba 
&& \wt\phi =\sum_{k=1}^K\left[x_{k}(t,\ta)e^{ik\ta}+ y_k^*(t,\ta)e^{-ik\ta}\right],
\lab{y0}\\
&& x_k=X_{k}(\ta) e^{-it\beta_k+t\lambda_k}, y_k=Y_{k}(\ta) e^{-it\beta_k+t\lambda_k}.
\lab{y1}
\ea
Here, $k$ is the Bloch angular momentum. $X_{k}(\ta)$ and $Y_{k}(\ta)$ are the cell-periodic functions, i.e., have the same periodicity as $\Phi(\ta)$, therefore their Fourier series are
\be
	 X_{k}(\ta)= \sum_l X_{k,l} e^{i lK\ta},~
	 Y_{k}(\ta)= \sum_l Y_{k,l} e^{i lK\ta}.
\lab{y9}
\ee

Substituting Eq.~\bref{y0} into Eq.~\bref{xs1} gives
\ba
	i(\p_t&&+\tfrac{1}{2}\kappa)\ket{{\bf Q}_k}
	=\pz
	\wh H_k\ket{{\bf Q}_k},~\ket{{\bf Q}_k}=\begin{bmatrix}x_k\\ y_k\end{bmatrix},\lab{ua}\\
	\wh H_k&&=	\begin{bmatrix}
		\wh D_k & -\gamma\Phi^{2} \\
		-\gamma\Phi^{*2} & \wh D_{-k}^{*}
	\end{bmatrix}, \ta\in[0,2\pi/K), \lab{ub}
	\\
	\wh D_{k}&&=\delta-2\gamma|\Phi|^2+\sum_{j\geqslant 1}\frac{d_j}{ j!} \left(k-i\p_\ta\right)^j,
\nn 
\ea
and then, applying Eq.~\bref{y1}, we find
\be
(\beta_{k}+i\lambda_{k}+i\tfrac{1}{2}\kappa)\ket{{\bf u}_{k}}=\pz\wh H_k\ket{{\bf u}_{k}},~
\ket{{\bf u}_k}=\begin{bmatrix}X_k\\ Y_k\end{bmatrix}.
\lab{zb2}
\ee
The above is Eq.~\bref{jb} in the main text.
One computational advantage of the Bloch formalism comes from the property that
\be
\beta_k=\beta_{k+K}, \lambda_k=\lambda_{k+K}, \ket{{\bf u}_{k}}=\ket{{\bf u}_{k+K}}e^{iK\ta},
\ee
and, therefore, $k$ can be restricted to a single Brillouin 
zone, $k=1,2,\dots,K$. The other is that the spectral bandwidth 
$\mu\in [-\mu_{\max},\mu_{\max}]$, see Eq.~\bref{e2}, is now achieved by taking $l\in [-l_{\max},l_{\max}]$ with $l_{\max}=\mu_{\max}/K$. For $K$ starting from 2 and above, this 
allows to work with the proportionally smaller number of modes when Eq.~\bref{zb2} is solved in the Fourier space.

\vspace{2mm}\noindent{\bf Geometrical phase and gauge invariance}\vspace{1mm}\\
\noindent 
The Zak phase formalism that needs to be developed for the soliton metacrystals
should  reflect on two features - (A)~nonlinear interaction between the phonon and crystal and (B) the discrete sampling of the states in the Brillouin zone. 
In order to make our derivations transparent, we first neglect (B) and, hence, 
take the limit of a large number of unit cells, so that $k$ is 
continuous in the Brillouin zone. $k$ is now assumed to vary adiabatically in time, $k=k(t)$.
The equation of motion, Eq.~\bref{ua}, for the phonon wave function, 
$\ket{{\bf Q}^{(\al)}_{k,n}}$,  is solved by the substitution~\cite{ber,zak2} \be
\ket{{\bf Q}^{(\al)}_{k,n}}= f_{k,n}^{(\al)} (t)\ket{ {\bf u}^{(\al)}_{k,n}},\lab{m1}\ee  
where~$f_{k,n}^{(\al)}$ is the function to be found and the phonons are assumed confined to the band $n$.
For $|\om_{\kk,n}|\gg |\lambda_{\kk,n}|,\kappa$,  the left eigenvector of $\pz H_k^{(\al)}$ is $\bra{{\bf u}^{(\al)}_{k,n}}\pz$ \cite{a4}. Substituting Eq.~\bref{m1} to Eq.~\bref{ub} and projecting  leads to
\begin{eqnarray}
&&i\frac{\text{d}f_{k,n}^{(\al)}}{f_{k,n}^{(\al)}}
\approx \beta_{\kk,n}\text{d}t-\cA^{(\al)}_{k,n}\text{d}k,\lab{ge}\\
	&&\cA^{(\al)}_{\kk,n}= 
	\frac{i}{N_n}\bra{ {\bf u}^{(\al)}_{\kk,n}}\pz\p_\kk \ket{{\bf u}^{(\al)}_{\kk,n}},~ N_n=\bra{ {\bf u}^{(\al)}_{k,n}}\pz\ket{ {\bf u}^{(\al)}_{k,n}},\nn
\end{eqnarray}
where $N_n$ is the normalization constant.

The $\beta_{k,n}$ and $\cA_{k,n}^{(\al)}$ terms in Eq.~\bref{ge} are the instantaneous dynamical and geometrical phases. Hence, the total geometric phase acquired during $k(t)$ passing the first Brillouin zone (BZ) is 
\be
\cz^{(\al)}_n=\int_{BZ}\cA^{(\al)}_{k,n}\text{d}k.
\lab{in}\ee  
Our $\cA_{\kk,n}^{(\alpha)}$ differs from the Berry connection emerging in the linear problems~\cite{ber,zak2,book}, by the Pauli matrix $\pz$ before $\p_\kk$, 
which has been previously reported in, e.g., photonic crystals with parametric gain~\cite{p1} and 
LC-circuits~\cite{p2,p3}.  The normalisation condition applied by us is $N_n=\pm 1$, see, e.g.,~\cite{a4,bog} for how similar norms appear in the theory of Bose condensates. $\pz$  also enters  the Berry curvature of the Bogolyubov excitations of Bose condensates~\cite{bog2}.

The phases of the Bloch functions are not 
uniquely defined and can be changed by the gauge transform, which, in the discrete BZ, takes the form
\be
{\bf u}^{(\al)}_{k,n}\to {\bf u}^{(\al)}_{k,n}e^{i\xi_k},~\xi_k=\xi_{k+K}.
\lab{gau}
\ee 
For the $k$-continuous, $\cz^{(\al)}_n$ remains invariant on applying Eq.~\bref{gau}~\cite{zak2}. When $k$ is sampled discretely, a variety of available numerical schemes to deal with $\p_k$ will not generally comply with the gauge invariance. Therefore, designing a practical algorithm to compute $\cz^{(\al)}_n$ is an important issue to consider.  

This brings us to the point (B) and we assume that $k$ is advancing in the unitary steps from $1$ to $K$, 
and each step takes the time d$t$ to make, then  Eq.~\bref{ge} 
updates as
\ba
&&i\frac{f_{k+1,n}^{(\al)}-f_{k,n}^{(\al)}}{f_{k,n}^{(\al)}}\approx\beta_{k,n}	\text{d}t
\lab{g3} \\
&&
-i\left(\bra{{\bf u}^{(\al)}_{k,n}}\pz\ket{{\bf u}^{(\al)}_{k+1,n}}-
\bra{{\bf u}^{(\al)}_{k,n}}\pz\ket{{\bf u}^{(\al)}_{k,n}}\right)/N_n,
\nn
\ea
and, therefore, the total geometric phase is 
\be
\cz_{n}^{(\al)}=-i\sum_{k=1}^{K}
\left(\bra{ {\bf u}^{(\al)}_{k,n}}\pz\ket{{\bf u}^{(\al)}_{k+1,n}}-N_n\right)/N_n.
\lab{ga1}
\ee
Checking of Eq.~\bref{ga1}  reveals the 
discretization scheme induced violation of the gauge invariance. Indeed, Eq.~\bref{ga1} is invariant on applying  
$({\bf u}^{(\al)}_{k,n},{\bf u}^{(\al)}_{k+1,n})\to 
({\bf u}^{(\al)}_{k,n}e^{i\xi_k},{\bf u}^{(\al)}_{k+1,n}e^{i\xi_k})$ and not relative to Eq.~\bref{gau}.

The recipe to restore the gauge invariance is given by the theory of polarization in crystalline solids~\cite{resta,van1}.
We first take the exponent from Eq.~\bref{ga1},
and then use  $e^x=1+x+\dots$, 
\begin{eqnarray}
\exp\{i\cz_{n}^{(\al)}\}&&=
\prod_{k=1}^{K}
\exp\left\{
\bra{ {\bf u}^{(\al)}_{k,n}}\pz\ket{{\bf u}^{(\al)}_{k+1,n}}\frac{1}{N_n}-1\right\}
\nn \\
&&\approx
\prod_{k=1}^{K}
\bra{ {\bf u}^{(\al)}_{k,n}}\pz\ket{{\bf u}^{(\al)}_{k+1,n}}\frac{1}{N_n}.
\lab{pr9}
\end{eqnarray}
Taking the logarithm gives the gauge invariant expression for  the geometric phase of the soliton crystals,
\begin{eqnarray}
&&\cz_{n}^{(\al)}=\text{Im}\ln \prod_{k=1}^{K}
\bra{ {\bf u}^{(\al)}_{k,n}}\pz\ket{{\bf u}^{(\al)}_{k+1,n}}\frac{1}{N_n} 
\lab{ga2}\\
&&=\text{Im}\ln \Big\{\frac{1}{N_n^{K}}\times
\bra{ {\bf u}^{(\al)}_{1,n}}\pz\ket{{\bf u}^{(\al)}_{2,n}}
\bra{ {\bf u}^{(\al)}_{2,n}}\pz\ket{{\bf u}^{(\al)}_{3,n}}\dots\nn\\
&&\dots\bra{ {\bf u}^{(\al)}_{K-1,n}}\pz\ket{{\bf u}^{(\al)}_{K,n}}
\bra{ {\bf u}^{(\al)}_{K,n}}\pz\ket{{\bf u}^{(\al)}_{1,n}e^{-iK\theta}}\Big\}.
\nn
\end{eqnarray}
'$\ln$' is defined by its principal value, such that
$\cz_{n}^{(\al)}\in [-2\pi,0)$.
For $K$ even, $N_n^K=1$ and the above coincides with Eq.~\bref{pro} in the main text.

\vspace{2mm}
\noindent{\bf Wannier formalism}\vspace{1mm}\\
The direct and inverse Bloch-to-Wannier transforms are defined as 
\be
\begin{split}
	& \ket{{\bf u}_{k,n}^{(\al)}}=
	\sum_{J}\ket{{\bf w}_{J,n}^{(\al)}}e^{-ik(\ta-\ta_J)}
	=\begin{bmatrix}X^{(\al)}_{k,n}\\ Y^{(\al)}_{k,n}\end{bmatrix},\\
	& \ket{{\bf w}_{J,n}^{(\al)}}=\frac{1}{K}
	\sum_{k} \ket{{\bf u}_{k,n}^{(\al)}}e^{ik(\ta-\ta_J)}
	=\begin{bmatrix}W^{(\al)}_{J,n}\\ V^{(\al)}_{J,n}\end{bmatrix},
\end{split}
\lab{b7}
\ee
where $\ket{{\bf w}_{J,n}^{(\al)}}$ are Wannier functions,
$\ta_J=2\pi(J-1)/K$ and 
$J\in\mathbb{Z}$ numbers the unit cells.
Our gauge choice for the Bloch functions is specified by the condition 
Im$X^{(\al)}_{k,n}=0$ at the point $\ta=3\pi/2K$.

We now recall Eq.~\bref{zb2} and, then, transform Eq.~\bref{in}, 
\ba
&&\cz_{n}^{(\al)}=i\int_{BZ} dk\bra{ {\bf u}^{(\al)}_{\kk,n}}\pz\p_\kk \ket{{\bf u}^{(\al)}_{\kk,n}}\lab{zr0}
\\&&
	=i\int_{BZ} dk\int_{\ta_J-\pi/K}^{\ta_J+\pi/K}\left(X^{(\al)*}_{k,n}\p_k X^{(\al)}_{k,n}-Y^{(\al)*}_{k,n}\p_k Y^{(\al)}_{k,n}\right)d\ta
	\nn\\
&&	=\frac{i}{K}\int_{BZ} dk\int_{\ta_J-\pi}^{\ta_J+\pi}\left(X^{(\al)*}_{k,n}\p_k X^{(\al)}_{k,n}-Y^{(\al)*}_{k,n}\p_k Y^{(\al)}_{k,n}\right)d\ta.
\nn
\ea
The last part of Eq.~\bref{zr0} uses the fact that the Bloch functions are cell periodic and, hence, the integration can be extended to the entire ring. 

If the ring is large, the problem can be well approximated by 
the straight-line infinite crystal lattice~\cite{zak2}
with the period  $2\pi/K$. Then, $\sum_k\to\int_{BZ}dk$ and 
expressing the Bloch functions via the Wannier ones
allows to take the $k$-derivative inside Eq.~\bref{zr0},
\ba
	\cz_{n}^{(\al)}&=&
	\sum_{J}\int_{\ta_{J}-\pi}^{\ta_{J}+\pi}
	\left[
	|W^{(\al)}_{J,n}|^2
	-
	|V^{(\al)}_{J,n}|^2
	\right](\ta-\ta_J)d\ta\nn\\
	&=&\sum_{J}\bra{ {\bf w}^{(\al)}_{\bj,n}}\pz(\ta-\ta_\bj)\ket{{\bf w}^{(\al)}_{\bj,n}}.
\lab{p1}
\ea
Wannier functions for all $J$  coincide if plotted vs $\ta-\ta_J$,
therefore,
Eq.~\bref{p1} simplifies to
\be
\cz_{n}^{(\al)}=	K\bra{ {\bf w}^{(\al)}_{J,n}}\pz(\ta-\ta_J)\ket{{\bf w}^{(\al)}_{J,n}},
\lab{p2}
\ee 
see Eq.~\bref{z0} in the main text.
Thus, the geometric phase of soliton crystals is determined by the coordinate of the effective centre of the phonon cloud in the unit cell, i.e., by the Wannier-centre, $\bra{ {\bf w}^{(\al)}_{J,n}}\pz(\ta-\ta_J)\ket{{\bf w}^{(\al)}_{J,n}}$. 
For the ring geometry with the relatively large number of unit cells, as we are dealing with, Eq.~\bref{p2} remains a good  approximation.

\vspace{3mm}\noindent{\bf Data availability}\\
The data supporting the findings of this study are available from authors on reasonable request.


\vspace{8mm}\noindent{\bf Acknowledgements}\\
We acknowledge the financial support received from  
EU Horizon 2020 Framework Programme (812818, MICROCOMB)
and UK EPSRC (2119373).

\end{document}